\documentclass[a4paper,12pt]{article}
\usepackage{a4wide,latexsym}
\usepackage{epsfig,graphicx,subfigure,amsmath}

\newcommand{\bea}{\begin{eqnarray}}
\newcommand{\eea}{\end{eqnarray}}
\newcommand{\be}{\begin{equation}}
\newcommand{\ee}{\end{equation}}
\newcommand{\pkt}{\; .}
\newcommand{\kma}{\; ,}
\newcommand{\eqn}[1]{(\ref{#1})}

\newcommand{\tr}{{\rm Tr}}

\newcommand{\calm}{{\cal M}}

\newcommand{\calf}{{\cal F}}
\newcommand{\cald}{{\cal D}}
\newcommand{\calr}{{\cal R}}
\newcommand{\calg}{{\cal G}}
\newcommand{\calb}{{\cal B}}
\newcommand{\caln}{{\cal N}}
\newcommand{\calc}{{\cal C}}
\newcommand{\cals}{{\cal S}}

\newcommand{\call}{{\cal L}}

\newcommand{\bfS}{{\bf S}}

\begin{document}

\begin{titlepage}

\begin{flushright}
\texttt{DO-TH-04/08}\\
July 2004
\end{flushright}

\vspace{20mm}
\begin{center}
{\Large \bf \sffamily The 2PI finite temperature effective potential of 
  the O(N) linear sigma model in 1+1 dimensions
  at next-to-leading order in 1/N}

\vspace{10mm}

{\large  J\"urgen Baacke\footnote{
    e-mail:~\texttt{baacke@physik.uni-dortmund.de}}} 
\hspace{.15cm} and \hspace{.15cm}
{\large Stefan Michalski\footnote{
    e-mail:~\texttt{stefan.michalski@uni-dortmund.de}} } 
\\\vspace{.5cm}
{\em Institut f\"ur Physik, Universit\"at Dortmund \\
  D--44221 Dortmund, Germany}
\\
\vspace{8mm}
\end{center}

\begin{abstract}
\noindent
We  study  the $O(N)$ linear sigma model in 1+1 dimensions
by using the 2PI formalism
of Cornwall, Jackiw and Tomboulis in order to evaluate
the effective potential at finite temperature. At next-to-leading
order in a $1/N$ expansion one has to include the sums
over ``necklace''' and generalized ``sunset'' diagrams.
 We find that --- in contrast to the Hartree approximation ---
there is no spontaneous symmetry breaking in this approximation, 
as to be expected for the exact theory.
The effective potential becomes convex throughout for all parameter sets
which include $N=4,10,100$, couplings $\lambda=0.1,0.5$
and temperatures between $0.3$ and $1$ (in arbitrary units). 
The Green's functions obtained
by solving the Schwinger-Dyson equations are enhanced in the
infrared region. 
We also compare the effective potential as a function of the
external field $\phi$ with those obtained in the 1PI and 2PPI expansions.
\end{abstract}

\end{titlepage}

\section{Introduction}
The $O(N)$ linear sigma model at finite temperature
has a long-standing history, in particular
as a basic model for a quantum field theory with spontaneous symmetry breaking
\cite{Kirzhnits:1975as,Coleman:1974jh,Dolan:1974qd,Bardeen:1983st}.
Early investigations beyond the classical level have been based on
including one-loop quantum and thermal corrections.
These studies have been centered around the discussion of the 
one-loop effective potential $V_{\rm eff}(\phi)$ where $\phi$ is the
mean value of the quantum field $\Phi$, in a sense being
defined more precisely by the effective action formalism,
summing up \emph{one-particle irreducible} (1PI) graphs.
A next class of approximations include bubble resummations, usually
obtained from the 1PI formalism by introducing an auxiliary field. 
Within such a framework
the effective potential has been computed to leading 
\cite{Coleman:1974jh,Dolan:1974qd,Schnitzer:1974ji} and 
next-to-leading order (NLO) of a $1/N$ expansion \cite{Root:1974zr}.
At leading order in  $1/N$ one does not find spontaneous symmetry breaking
in $1+1$ dimensions, while in $3+1$ dimensions the effective
potential is flat for $|\vec \phi| < v$ \cite{Bardeen:1983st}
\footnote{This follows from an end point extremum of 
  $\Gamma(\phi,M^2)$ at $M^2=0$.}. 
At next-to leading order
there is no spontaneous symmetry breaking in $1+1$ dimensions either.
These results correctly reproduce a general property of theories
in $1+1$ dimensions, where spontaneous symmetry breaking 
is prevented by the non-existence of Goldstone 
bosons~\cite{Coleman:1973ci}.

Another approximation including some nonleading terms of
a $1/N$ expansion is the Hartree approximation. It is usually
motivated by a self-consistency of one-loop quantum corrections.
It has been studied in $3+1$ dimensions by various authors,
in thermal equilibrium \cite{Amelino-Camelia:1993nc,Amelino-Camelia:1997dd,
Nemoto:1999qf, Lenaghan:1999si,Verschelde:2000ta},
and out-of-equilibrium \cite{Baacke:2001zt}.
The model with spontaneous symmetry
breaking is found to have a phase transition of first order towards 
the symmetric phase at high temperature. In $1+1$ dimensions
the Hartree approximation displays spontaneous symmetry breaking and symmetry 
restoration as in $3+1$ dimensions, an obviously unphysical
feature. 

If one wants to go beyond the large-$N$ and Hartree approximations
there is a variety of choices.
One of those is the use of the \emph{two-particle irreducible} (2PI) 
formalism of
Cornwall, Jackiw and Tomboulis (CJT) \cite{Cornwall:1974vz}
within which one may include higher loop 
corrections, higher orders in $1/N$ etc., as in a 1PI expansion.
But furthermore it becomes necessary to solve Schwinger-Dyson equations 
for the propagators in order to sum up certain classes of Feynman graphs. 
Apart from the fact that this may be technically
demanding, there still is the obstacle that for $3+1$ dimensions
renormalization has been discussed up to now only for vanishing 
external fields~\cite{vanHees:2001ik,vanHees:2001pf,vanHees:2002bv,
Blaizot:2003an};  this obstructs the discussion of 
spontaneous symmetry breaking, which here is our main interest.
As here we discuss the model in $1+1$ dimensions, the problem
of renormalization does not arise and explicit numerical computations 
can be performed. This is indeed the main subject of this
publication.

A technically less demanding approach is the so-called 
\emph{two-particle point-irreducible} (2PPI) resummation introduced
by Coppens and Verschelde \cite{Verschelde:1992bs,Coppens:1993zc}. 
Here, instead of treating
the Green's functions as variational parameters, 
one just introduces variational
masses like in the Hartree approximation. This implies that the resummation
is only over local insertions, the 2-particle point-reducible
graphs, i.e., graphs that fall apart if two lines meeting at the same
point (the 2PPR point) are cut. 
This approach is identical to the Hartree approximation
if only one-loop 2PPI graphs are included.  
For this formalism renormalization has been fully 
discussed \cite{Verschelde:2000dz}; 
symmetry restoration at finite temperature has been
investigated in $3+1$ dimensions in an approximation including
the sunset diagram, for the case $N=1$ \cite{Smet:2001un} and 
for the $O(N)$ model with arbitrary $N$ \cite{Baacke:2002pi}. 


The  2PI formalism has the advantage that at a given order
of the loop expansion it resums a larger
class of Feynman diagrams than the 2PPI or Root's auxiliary
field scheme.  From a variational
point of view it allows for a more ``flexible'' propagator whose
self-energy insertions are nonlocal in space-time and thereby 
become momentum dependent. 

It is the purpose of the present
work to study the 2PI formalism at next-to-leading order
of the $1/N$ expansion, an approximation usually denoted as 2PI-NLO 
for short.
For the $O(N)$ model the next-to-leading order 1PI diagrams 
have been identified in the
work of Root \cite{Root:1974zr} using an auxiliary field method. 
They are depicted in 
Figs.~\ref{fig:pearls} and \ref{fig:chains}
in a way that graphically differs from Root's presentation.
The 2PI-NLO diagrams are obtained by omitting those which are
two particle reducible, they have been discussed 
in Refs. \cite{Berges:2001fi,Aarts:2002dj}.
The sums over all ``necklace'' (cf. Fig.\ref{fig:pearls}) and 
``generalized sunset'' (cf. Fig.~\ref{fig:generalsunset})
diagrams can be done in closed form.  

We have evaluated the effective potential at finite temperature in  the
2PI-NLO scheme by solving the Schwinger-Dyson gap equations. For comparison we 
will also show results obtained in the other schemes introduced above.
 

The consideration of $\Phi^4$ theory in $1+1$ dimensions has recently 
received some interest in the context of nonequilibrium dynamics
\cite{Berges:2001fi,Berges:2000ur,Aarts:2001yn,Cooper:2002qd,Baacke:2002ee}.
A comparison \cite{Cooper:2002qd,Baacke:2003qh}
of various approximations for the $O(1)$ model, i.e., a model with
a simple double-well potential, has displayed some still not well-understood
features. The $O(1)$ model has thermal kink transitions 
\cite{Grigoriev:1988bd,Alford:1991qg,Alexander:1993ha,Salle:2003ju}
that lead to a symmetric phase in the exact theory.
These are out of reach within
a perturbative treatment in thermal equilibrium, 
whereas they may play (or seem to play) 
a non-perturbative r\^ole in nonequilibrium physics 
even in the approximations considered
here. Our present calculations are not directly relevant to this
case. However, the comparison of the various schemes for $N>1$ should
be of interest as well, both in equilibrium and out of equilibrium.
Here we provide the equilibrium computations, in expectation of their
nonequilibrium counterparts.

The plan of the paper is as follows: 
 in Section~\ref{sec:basics} we present the general formulation 
of the model and
of the 2PI formalism at NLO-$1/N$. We also present all the 
formulae used in computing the finite-temperature effective
potential in the NLO-$1/N$ approximation. This discussion includes
the Hartree approximation. The analogous derivations for the 2PPI and
1PI (Root) scheme 
are presented in Appendix~\ref{sec:2PPI} and~\ref{sec:1PI1/N}. 
In Section~\ref{sec:results} we discuss our numerical results.
We end with some conclusions and an outlook in Section~\ref{sec:conclusions}.

\section{Basic equations}
\label{sec:basics}
The Lagrange density of the linear sigma
is given by
\be
\call =\frac{1}{2}\,\partial_\mu\vec\Phi\cdot\partial^\mu\vec\Phi
-\frac{\lambda}{4N}\left(\vec\Phi^2-Nv^2\right)^2
\pkt
\ee
The 2PI effective action formalism
introduces two variational functions: an external field
$\phi(x)$ and a Green's function $\calg(x,x')$, 
which arise as Legendre conjugate variables related to a
local source term $J(x)\Phi(x)$ and a bilocal source
term $K(x,x')\Phi(x)\Phi(x')$.
Assuming that the expectation value of $\vec \Phi^2$ will scale
as $N$ we introduce the external field $\phi$ with a scale
factor $\sqrt{N}$, i.e., $\langle \Phi \rangle = \sqrt{N} \phi$.
 We then obtain the classical potential
\be
V_{\rm class}(\phi)=\frac{\lambda}{4N}\left(N\phi^2-Nv^2\right)^2
=N\frac{\lambda}{4}\left(\phi^2-v^2\right)^2
\pkt
\ee

The one-loop part of the effective action is given by
\be
\Gamma_{\rm 1-loop}[\phi,\mathcal{G}]=
  \frac{i}{2}\tr \ln \calg^{-1}/\calg_0^{-1}
  + \frac{1}{2}\tr \left[ i\cald^{-1} \calg-1\right]
\pkt\ee
Here $\calg$, $\calg_0$ and $\cald$ are matrices. If we separate
the fields in a basis parallel and orthogonal to the direction
$\hat n$ of the classical field $(\vec \Phi)_{\rm cl} = \sqrt{N}\vec n \phi$
these matrices become diagonal, with one entry for the parallel
component and $(N-1)$ identical entries for the orthogonal ones. We denote them
as $\sigma$ and $\pi$ in reference to the linear sigma model, which
presents one of the possible applications.
With this notation we have
\be
i\cald_j^{-1}=-\partial_\mu\partial^\mu - \lambda (f_j \phi^2- v^2)
\pkt \ee
This is the Klein-Gordon operator in the external field; $f_j=3$
for $j=\sigma$ and $f_j=1$ for $j=\pi$. We have normalized
the Green's functions $\calg_j$ with respect to  free Green's functions
\be
\calg_{j,0}^{-1}=-\partial_\mu\partial^\mu - m_{j,0}^2
\pkt
\ee
The higher terms in the effective action are
all two-particle irreducible vacuum Feynman graphs with external lines 
$\sqrt{N}\phi(x)$ and with Green's function $\calg$ as
internal lines. 

In the following we will consider the system at finite 
temperature. So we perform the Wick rotation and replace the
integration  over the Euclidean $p^0$ by the summation over the
Matsubara frequencies $p^0_n=\omega_n=2n \pi k_B T$. 
Furthermore we consider the effective potential instead of
the effective action. 
Then the one-loop term for one species, $j=\sigma$ or
$j=\pi$ takes the form
\bea
V_{\rm 1-loop,j}&=&\frac{1}{2}T \sum_{m=-\infty}^\infty\int_{-\infty}^\infty
\frac{dq}{2\pi} \left \{ \ln \frac{q^2+\omega_m^2
+m_{j,0}^2+\Sigma_j(\omega_m,q)}{q^2+\omega_m^2+m_{j,0}^2}\right.
\\\nonumber
&& \left.- \left[m_{j,0}^2+\Sigma_j(\omega_m,q)-\lambda(f_j\phi^2-v^2)\right] 
\calg_j(\omega_m,q) \right \}
\pkt\eea
Here  
\be
\calg_{j,0}^{-1}(\omega_n,p)=p^2+\omega_n^2+m_{j,0}^2
\ee
is the 
reference Green's function, and $\Sigma_j(\omega_n,p)$ is defined
as 
\be
\Sigma_j(\omega_n,p)=\calg_j^{-1}(\omega_n,p)
-\calg_{j,0}^{-1}(\omega_n,p)
\pkt\ee
The choice of $m_{0j}$ only introduces an additive, though temperature
dependent, constant to the effective potential. An obvious choice would
be to use the bare physical masses, but this would imply $m_{\pi 0}^2=0$,
leading immediately to infrared singularities. As we anticipate that 
the symmetry may not be broken in the NLO-$1/N$ approximation, we prefer a
symmetric choice: $m_{j0}^2=\lambda v^2$ for both $j=\sigma$ and
$j=\pi$.

The one-loop potential is divergent. The asymptotic behavior of the integrand
is
\be
\left\{\cdots\right\}\simeq-\frac{m_{j0}^2-\lambda(f_j\phi^2-v^2)}
{q^2+\omega_n^2+m_{j,0}^2}
\pkt
\ee
So the integral of the one-loop effective action can be 
regulated by subtracting this term and by adding it in regularized
form. We have
\bea
\nonumber
V_{\rm 1-loop,j}&=&\frac{1}{2}T \sum_{m=-\infty}^\infty\int_{-\infty}^\infty
\frac{dq}{2\pi} \left \{ \ln \frac{q^2+\omega_m^2
+m_{j,0}^2+\Sigma_j(\omega_m,q)}{q^2+\omega_m^2+m_{j,0}^2}\right.
\\\nonumber
&& \left.- \left[m_{j,0}^2+\Sigma_j(\omega_m,q)-\lambda(f_j\phi^2-v^2)\right] 
\calg_j(\omega_m,q) \right.
\\\nonumber
&&\left.+\frac{m_{j0}^2-\lambda(f_j\phi^2-v^2)}
{q^2+\omega_n^2+m_{j,0}^2}\right\}
\\
&&-\left[m_{j0}^2-\lambda(f_j\phi^2-v^2)\right]\calb_{j0}
\eea
where $\calb_j$ are the bubble diagrams which we regularize dimensionally as
\bea \label{bubble_reg}
\nonumber
\calb_{j,0}&=&T\sum_{m=-\infty}^\infty\int\frac{dq^{(1-\epsilon)}}{(2\pi)
^{(1-\epsilon)}}\calg_{j,0}(\omega_m,q)
\\
&=&\frac{1}{4\pi}\left[\frac{2}{\epsilon}-\gamma_E+
\ln 4\pi -\ln\frac{m_{j,0}^2}{\mu_j^2}\right]+
\int\frac{dq}{2\pi\omega}\,\frac{1}{\exp\beta\sqrt{q^2+m_{j,0}^2}-1}
\pkt
\eea
We define  their finite part $\calb_{j,0,\rm fin}$ using 
the $\overline{\rm MS}$ prescription,
leaving out the terms $2/\epsilon-\gamma_E+\ln(4\pi)$.
Then the renormalized bubble diagrams become
\be
\calb_j=T\sum_{m=-\infty}^\infty\int\frac{dp}{2\pi}
\left[\calg_j(\omega_n,p)-\calg_{j,0}(\omega_n,p)\right]
+\calb_{j,0,\rm fin}
\pkt
\ee 
with
\be
\calb_{j,0,\rm fin}=\frac{-1}{4\pi} \ln\frac{m_{j,0}^2}{\mu_j^2}
+\int\frac{dq}{2\pi\omega}\,\frac{1}{\exp\beta\sqrt{q^2+m_{j,0}^2}-1}
\pkt\ee
For the renormalization scale we choose $\mu_{j}^2=m_{j0}^2=\lambda v^2$.
In $1+1$ dimensions the theory is renormalized if the Hamiltonian
is normal-ordered. This is not a unique prescription as the 
normal-ordering may be done with respect to different 
masses of the bare quanta \cite{Coleman:1974bu,Chang:1976ek}.
A shift in these masses introduces a redefinition of the
vacuum expectation value $v$, or of the mass parameter in the symmetric
theory, as the change in normal ordering of the $\phi^4$ term introduces
a term quadratic in the fields. In the present context we have to take
the renormalization scale identical for the $\sigma$ and $\pi$ modes as
otherwise we break the symmetry explicitly. Our convention corresponds 
to normal-ordering with respect to bare quanta of mass $\lambda v^2$. 

With these preliminaries the total one-loop effective potential is given by
\be
V_{{\rm 1-loop}}=V_{{\rm 1-loop},\sigma}+(N-1)V_{{\rm 1-loop},\pi}
\pkt
\ee

The first nontrivial term is the double-bubble diagram.
As we consider here leading order and next-to-leading
order contributions we have to calculate this diagram
with the exact combinatorial factors for $O(N)$. It takes the form
\bea
\nonumber
V_{\rm db}&=&\frac{\lambda}{4N}
\left\{ [\mathcal{B}_\sigma + (N-1) \mathcal{B}_\pi]^2
  + 2\, [\mathcal{B}_\sigma + (N-1) \mathcal{B}_\pi]
\right\} \\
\label{eq:db}
&=& \frac{\lambda}{4N} \left[(N^2-1)\calb_\pi^2+
  2(N-1)\calb_\sigma\calb_\pi+3\calb_\sigma^2\right]
\pkt
\eea
If only these two contributions are considered, we obtain the
Hartree approximation, or daisy and super-daisy resummation.
In writing down this contribution we have replaced the divergent
bubbles subdiagrams by the finite ones. This requires  mass
and vacuum energy counterterms which here are fixed by the
minimal subtraction prescription which eventually may be replaced by
a precise renormalization condition.

From 
\be
V_{\rm{Hartree}}=
V_{\rm 1-loop}+V_{\rm db}
\ee
we obtain the gap equations
\begin{subequations}
\label{eq:gapHartree}
\bea
\Sigma^H_\sigma(\omega_n,p)&=&-m_{\sigma 0}^2+\lambda(3 \phi^2-v^2)
+\frac{\lambda}{N}\left[3\calb_\sigma + (N-1)\calb_\pi \right]
\\
\Sigma^H_\pi(\omega_n,p)&=&-m_{\pi 0}^2+\lambda(\phi^2-v^2)+
\frac{\lambda}{N}\left[\calb_\sigma + (N+1)\calb_\pi\right]
\eea
\end{subequations}
When taking the functional derivative with respect
to $\calg_\pi$ we will, here and below,  consider each pion ``individually'', 
disregarding factors $(N-1)$ which would cancel in the
gap equation. In this way the second equation is obtained directly,
and  the sigma and ``each'' pion field are treated on the same footing.

The large-$N$ limit is obtained by omitting the sigma contribution entirely.
Then the first of the gap equations is omitted as well, and the second one
only contains the pion bubble with $N/(N-1)$ replaced by 
unity.

\begin{figure}[htbp]
  \centering
  \epsfig{file=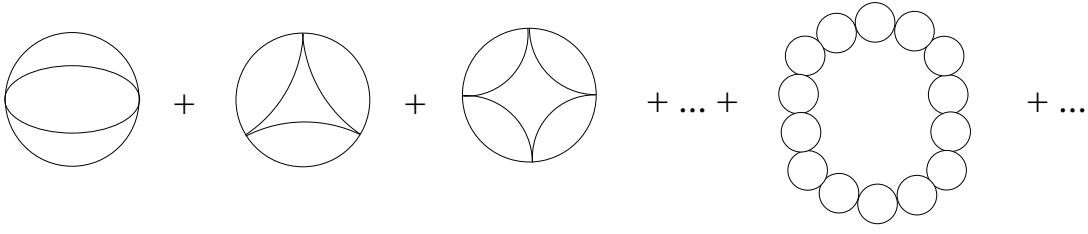,width=.9\textwidth}
  \caption{Resummation of necklace graphs. }
  \label{fig:pearls}
\end{figure}

The condition for a nontrivial extremum with respect to $\phi$ becomes
\be \label{defRh}
\calr(\phi)=\frac{1}{N\phi}\frac{\partial V_{\rm{eff}}^H}{\partial \phi}
=\lambda(\phi^2+\calb-v^2)=0
\ee
with the compound  bubble diagram
\be
\calb=\frac{\partial\left[3 V_{\rm{1-loop},\sigma}+
(N-1)V_{\rm{1-loop},\pi}
\right]}{\partial \phi}=
\frac{1}{N}\left[3\calb_\sigma+(N-1)\calb_\pi\right]
\pkt
\ee
Again in the large-$N$ limit the sigma contribution is neglected and
the prefactor $N/(N-1)$ of the pion bubble is replaced by unity.
To next-to-leading order in a $1/N$ expansion one has, in the absence
of external fields, to sum over all diagrams which are traces
of powers of the fish diagram, the so-called necklace diagrams, shown
in Fig. \ref{fig:pearls}.
In order to go beyond the leading order we define
several diagrams: 
We denote the fish diagrams  by $\calf_j(\omega_n,p)$; explicitly
\be \label{defFj}
\calf_j(\omega_n,p)=T\sum_{m=-\infty}^\infty\int\frac{dq}{2\pi}\
\calg_j(\omega_m,q) \calg_j(\omega_n-\omega_m,p-q)
\ee
and denote by $\calf(\omega_n,p)$ the ``average'' fish
\be\label{defF}
\calf(\omega_n,p)= \frac{1}{N}
\left[(N-1) \calf_\pi(\omega_n,p)+\calf_\sigma(\omega_n,p)\right]
\pkt
\ee
 The sum over the necklace without external fields, see Fig.~\ref{fig:pearls},
is then given by
\be \label{defN}
\caln=\frac{1}{2}T\sum_{m=-\infty}^\infty\int\frac{dq}{2\pi}
\left\{\ln [1+\lambda \calf (\omega_m,q)]-\lambda \calf(\omega_m,q)\right\}
\pkt
\ee
The subtraction of $\lambda \calf(\omega_m,p)$ accounts for the fact that we
have already included the double-bubble diagrams.
In the symmetric theory where
the $\sigma$ and $\pi$ contributions are equal, the
factor $N$ cancels, so this expression is of order $N^0$ which is NLO.

\begin{figure}[htbp]
  \centering
  \epsfig{file=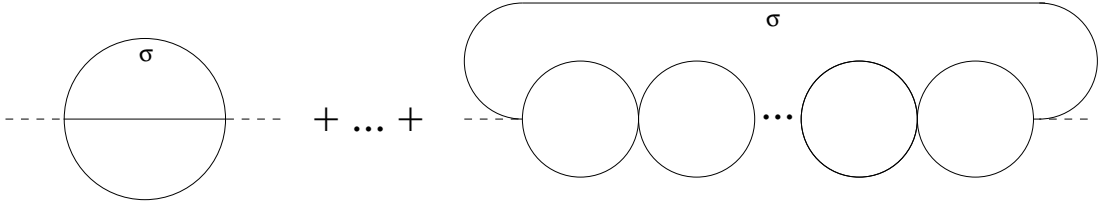,width=.9\textwidth}
  \caption{Generalized sunset graphs.}
  \label{fig:generalsunset}
\end{figure}

The same order in $1/N$ is achieved by replacing 
one of the $\sigma$ lines of the necklace by $N\phi^2$. 
We will call this class of graphs 
``generalized sunsets'', see Fig.~\ref{fig:generalsunset} for a graphical
depiction. 
We introduce the derivative
\be\label{defdNdG}
\frac{\delta \caln}{\delta \calg_j}
(\omega_n,p)=\frac{\lambda}{N} T\sum_{m=-\infty}^\infty\int\frac{dq}{2\pi}
\frac{-\lambda\calf(\omega_m,q)}{1+\lambda\calf(\omega_m,q)}\
\calg_j(\omega_n-\omega_m,p-q)
\pkt
\ee
Note that the factor $\lambda/N$ arises from the derivative of the
expression $\lambda\calf(\omega_n,p)$ with respect to $\calg_j$. 
In terms of $\delta \caln/\delta \calg_\sigma$ we
 obtain for the generalized sunset diagram, or 
sum over necklace diagrams with
two external lines
\be\label{defS}
\cals=N\phi^2\frac{\delta \caln}{\delta \calg_\sigma}(0,0)
=\lambda\phi^2 T\sum_{m=-\infty}^\infty\int\frac{dq}{2\pi}
\frac{-\lambda\calf(\omega_m,q)}{1+\lambda\calf(\omega_m,q)}
\calg_j(\omega_m,q)
\pkt\ee
The functional derivative contains a factor
$1/N$ so that this expression is of order $N^0$.
With these definitions the 2PI effective potential up to NLO becomes
\be
V_{\rm eff}=V_{\rm class}+V_{\rm 1-loop}+V_{\rm db}
 + \cals + \caln
\ee

In order to write down the gap equations we need to introduce
a further functional derivative
\be
\frac{\delta\cals}{\delta 
\calg_j}(\omega_n,p)=
N\phi^2\frac{\delta}{\delta \calg_j (\omega_n,p)}
\left[\frac{\delta \caln}{\delta \calg_\sigma}(0,0)\right]
\pkt
\ee
Explicitly it is given by
\bea
\nonumber
\frac{\delta\cals}{\delta 
\calg_j}(\omega_n,p)=\frac{\lambda^2}{N}\phi^2\,T\int \frac{dq}{2\pi}
\frac{-1}{[1+\lambda\calf(\omega_m,q)]^2}\calg_\sigma(\omega_m,q)
\calg_j(\omega_n-\omega_m,p-q)
\\
+\lambda\phi^2\delta_{j\sigma}\frac{-\lambda\calf(\omega_n,p)}
{1+\lambda\calf(\omega_n,p)}
\pkt
\eea
With these definitions the gap equations become
\begin{subequations}
\label{eq:gap2PI}
\bea\nonumber
\Sigma_\sigma(\omega_n,p)&=&-m_{\sigma 0}^2+\lambda(3\phi^2-v^2)
+\frac{\lambda}{N}\left[(N-1)\calb_\pi+3\calb_\sigma\right]
\\ \label{gap1}
&&+\frac{\delta \caln}{\delta \calg_\sigma}(\omega_n,p)
+\frac{\delta \cals}{\delta \calg_\sigma}(\omega_n,p)
\\\nonumber
\Sigma_\pi(\omega_n,p)&=&-m_{\pi 0}^2+\lambda(\phi^2-v^2)
+\frac{\lambda}{N}\left[(N+1)\calb_\pi+\calb_\sigma\right]
\\\label{gap2}
&&+\frac{\delta \caln}{\delta \calg_\pi}(\omega_n,p)
+\frac{\delta \cals}{\delta \calg_\pi}(\omega_n,p)
\eea
\end{subequations}
All expressions in these equations have been given explicitly above.
Finally for the partial derivative of $V_{\rm eff}$ with respect 
to $\phi$ we find
\be\label{defR}
\calr(\phi)=\frac{1}{N\phi}\frac{\partial V_{\rm{eff}}}{\partial \phi}
=\lambda(\phi^2+\calb-v^2)+2\lambda \frac{\delta \caln}{\delta \calg_\sigma}
(0,0)
\ee

We have performed the computations for the 2PPI scheme of Verschelde
and Coppens as well, including necklace and generalized sunset diagrams.
The formulae and their derivation are rather similar to those
of the 2PI scheme. They are given in Appendix~\ref{sec:2PPI}.


\section{Numerics and Results}
\label{sec:results}
The equations of the previous section are easily incorporated
into a computer code; we solve the gap equations by iteration.
The range of momenta and Matsubara frequencies was restricted
to values smaller than $15-20$, far above the relevant mass
scales. All numerical integrals and summations are ultraviolet
finite by the regularization presented in the previous section.

For the first step at fixed $N, v,\lambda, T$ and $\phi$ 
we start with $\calg_j=\calg_{j0}$. The iteration is 
monitored by the values of the average bubble diagram
\be
\overline{\calb} = \left[(N-1)\calb_\pi+\calb_\sigma\right]/N
\pkt\ee
If the relative change of this global quantity has become
smaller than $10^{-8}$ the iteration is considered to
have reached the solution.

For temperatures of the order of $v$ this procedure converges
well. Problems arise if the temperature is low, typically less then
$v/5$ and for $\phi$ typically less than $ v/2$. In these regions
the pion propagator becomes large at small $(\omega_n,p)$ obviously
due to a pion pole approaching $p^2=0$ from negative (Minkowskian)
$p^2$.  This situation is close to an infrared divergence,
and small changes of the pion Green's function result in
large changes of the various integrals. 
One can extend the domain of convergence by using an underrelaxation,
defining the $n$th step of the iteration of Eqs. \eqref{eq:gap2PI} by 
taking the result  $\Sigma^{(n-1)}_j$ of the previous step multiplied 
by a factor of $\alpha$ plus the right hand
sides of these equations multiplied by $(1-\alpha)$.
We have chosen $\alpha$ typically between $.5$ and $.95$.

In order to be sure of the consistency of our search for extrema,
both in the analytic formulas and in the numerical computations
 we have not only computed the results for 
$\partial V_{\rm eff}[\phi,\bar \calg_j]/\partial \phi$ 
but also the potential $V_{\rm eff}[\phi,\bar \calg_j]$ itself. 
Here in both cases
the Green's functions $\calg_j$ were the solutions
$\bar \calg_j$ of the
gap equations; so both $V_{\rm eff}$ and
$\partial V_{\rm eff}/\partial \phi$ are evaluated at 
$\delta V_{\rm eff}/\delta \calg =0$ and the extrema
of the potential  $V_{\rm eff}[\phi,\bar \calg_j]$ should
coincide with the zeros of
$\partial V_{\rm eff}[\phi,\bar \calg_j]/\partial \phi$. 
For the computations in the 2PI scheme we do not find any
nontrivial minima, so all we check here is that $\calr(\phi)$ has no
zeros. For the Hartree approximation and for the 2PPI scheme,
the extrema of the potential and the zeros of $\calr(\phi)$ agree
within the avalailable accuracy.

\begin{figure}[htbp]
  \centering
  \epsfig{file=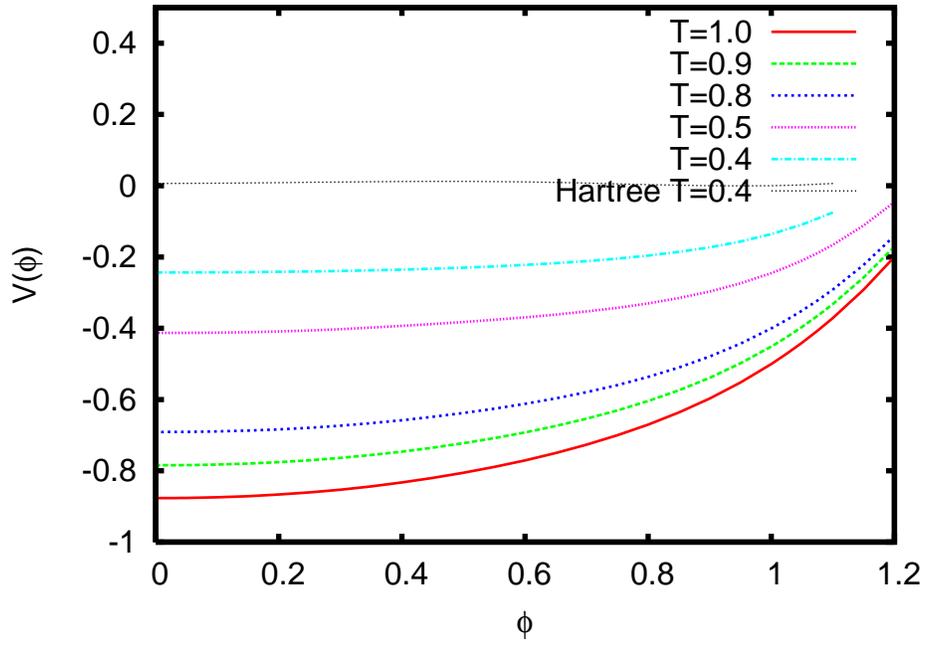}
  \caption{2PI effective potential at NLO-$1/N$ for $N=4$ and $\lambda=0.5$.}
  \label{fig:2PI_lambda0.5}
\end{figure}

\begin{figure}[htbp]
  \centering
  \epsfig{file=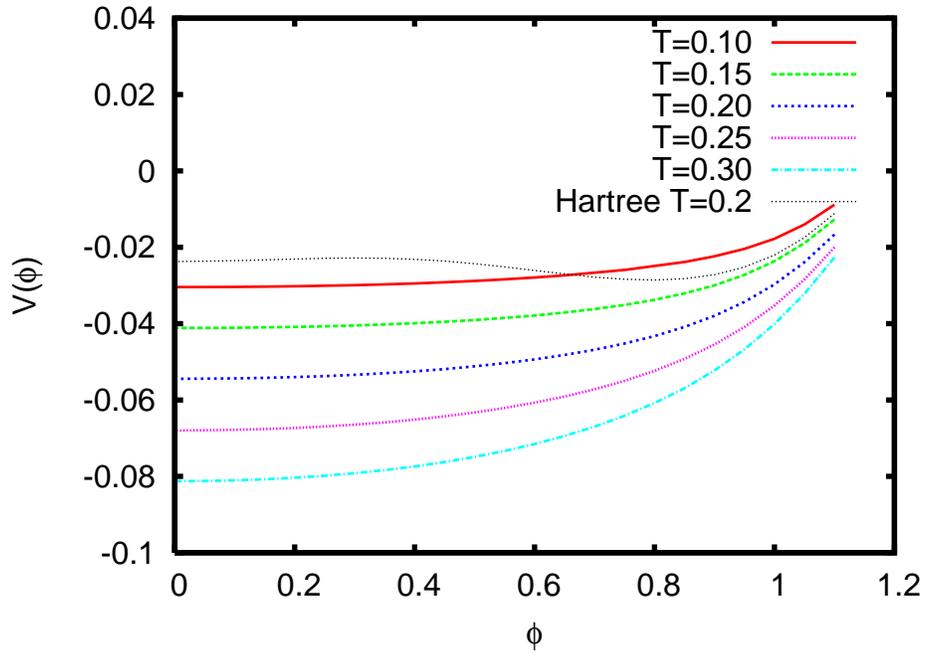}
  \caption{2PI effective potential at NLO-$1/N$ for $N=4$ and $\lambda=0.1$.}
  \label{fig:2PI_lambda0.1}
\end{figure}

In Figs.~\ref{fig:2PI_lambda0.5} and~\ref{fig:2PI_lambda0.1} we show 
the temperature dependence of the
2PI effective potential, for $N=4$ and $\lambda=0.5$ and
$\lambda=0.1$. The effective potential shows no sign
of an inflection point. We also plot the Hartree result for
one of the temperatures, in the neighborhood of the critical
temperature (of the Hartree approximation). 

\begin{figure}[htbp]
  \centering
  \subfigure[$N=4$]{
    \epsfig{file=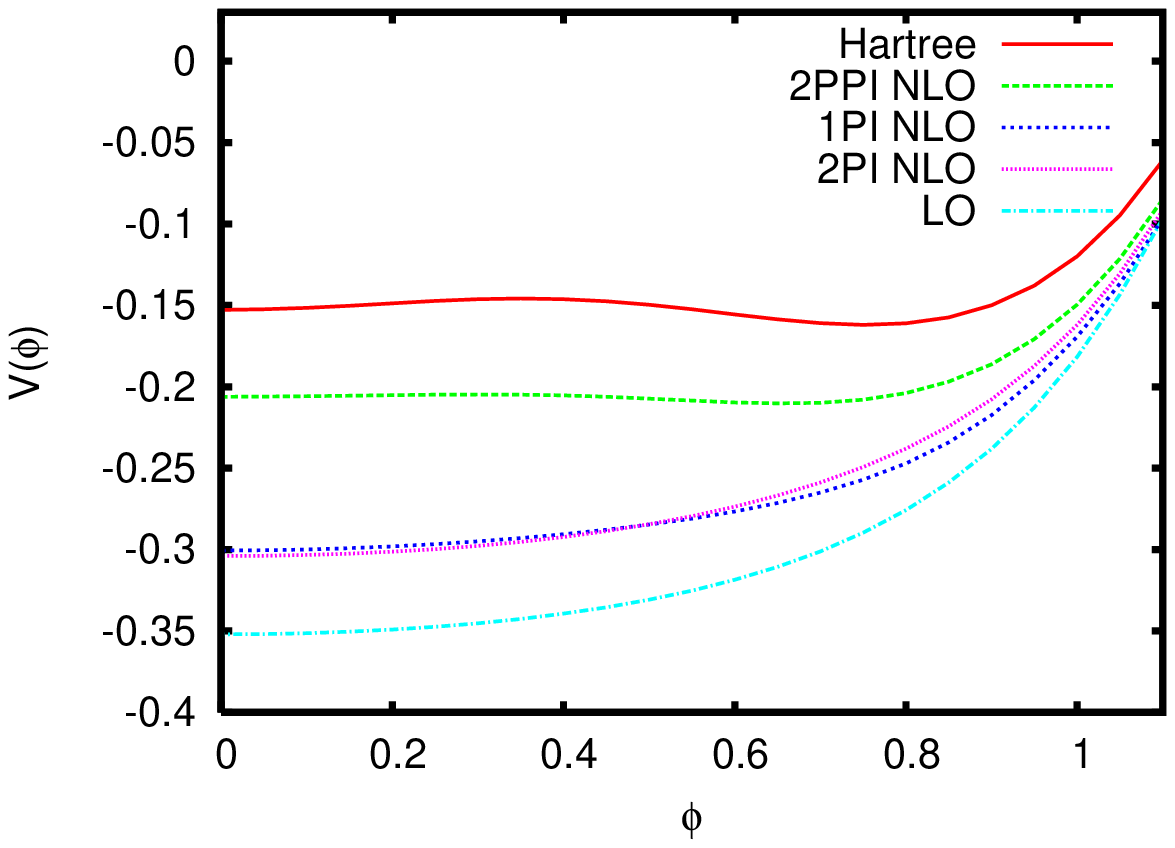,width=.47\textwidth}
  }
  \subfigure[$N=10$]{
    \epsfig{file=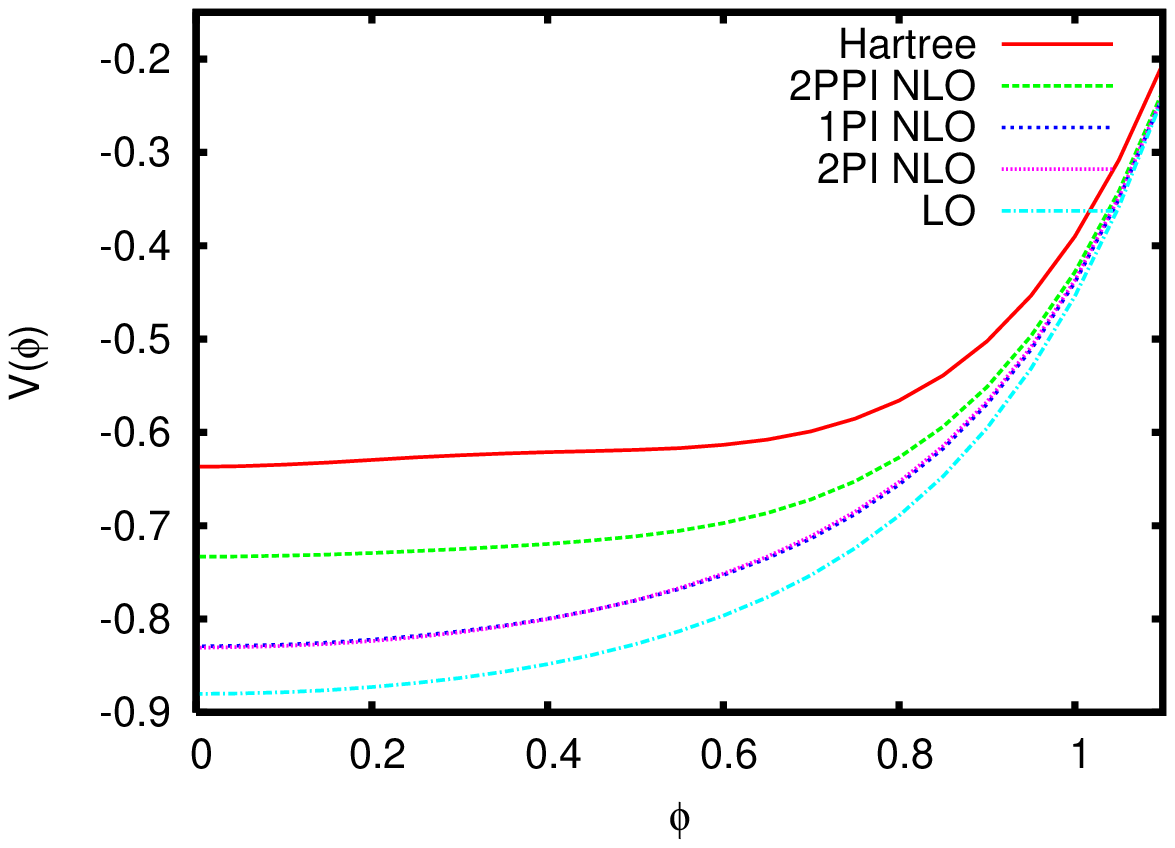,width=.47\textwidth}
  }
  \subfigure[$N=100$]{
    \epsfig{file=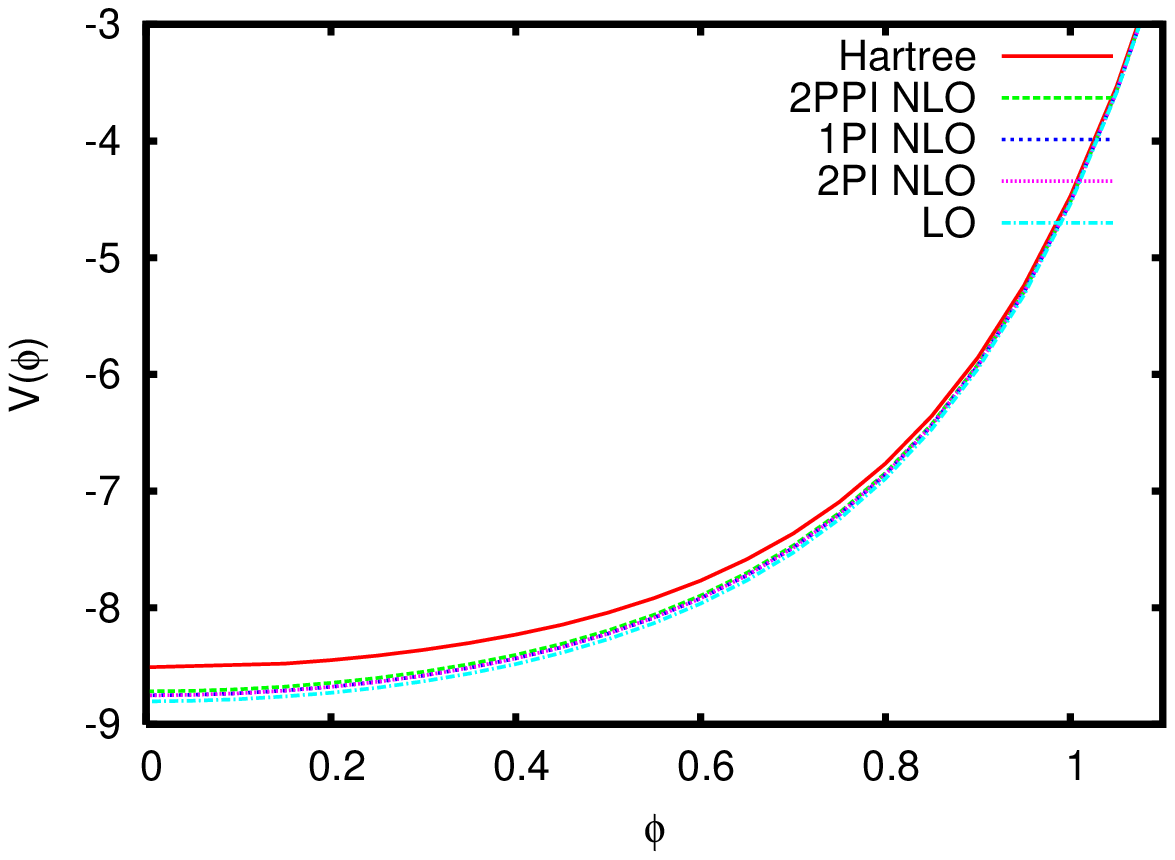,width=.47\textwidth}
  }
  \caption{Comparison of the effective potentials obtaind in the 
Hartree, 2PPI-NLO, 1PI-NLO, 2PI-NLO and leading order
large-$N$ approximations for $\lambda=0.5$, $T=0.5$  and 
    different values of $N$.}
  \label{fig:H_2ppi_2pi}
\end{figure}

In Fig.~\ref{fig:H_2ppi_2pi} we show the comparison between the Hartree,
the 2PPI-NLO and  Root's NLO approximation (denoted as
1PI-NLO for short, cf. Appendix~\ref{sec:1PI1/N}) and the convergence
towards the leading order large-$N$ results. These results are for
$v=1$, $\lambda=0.5$ and $T=0.5$. As one sees, the
effective potential in the 
2PPI approximation is always in between those of the Hartree and 2PI-NLO
approximation. The 1PI-NLO effective potential is close
to the 2PI results. The latter two approximations differ only by terms
of next-to-next-to-leading order (NNLO); it is surprising, nevertheless,
that these terms are small already at moderate values of
$N$. They are important for the evolution out of
equilibrium, as they include scattering of quantum fluctuations. 
As for finite $N$ one expects differences to
the large-$N$ result, the deviations of the
various approximations from this limit
do not a priori establish any ``ranking''; 
however, the spontaneous symmetry breaking
displayed by the Hartree and 2PPI-NLO 
approximations is certainly unphysical.

\begin{figure}[htbp]
  \centering
  \epsfig{file=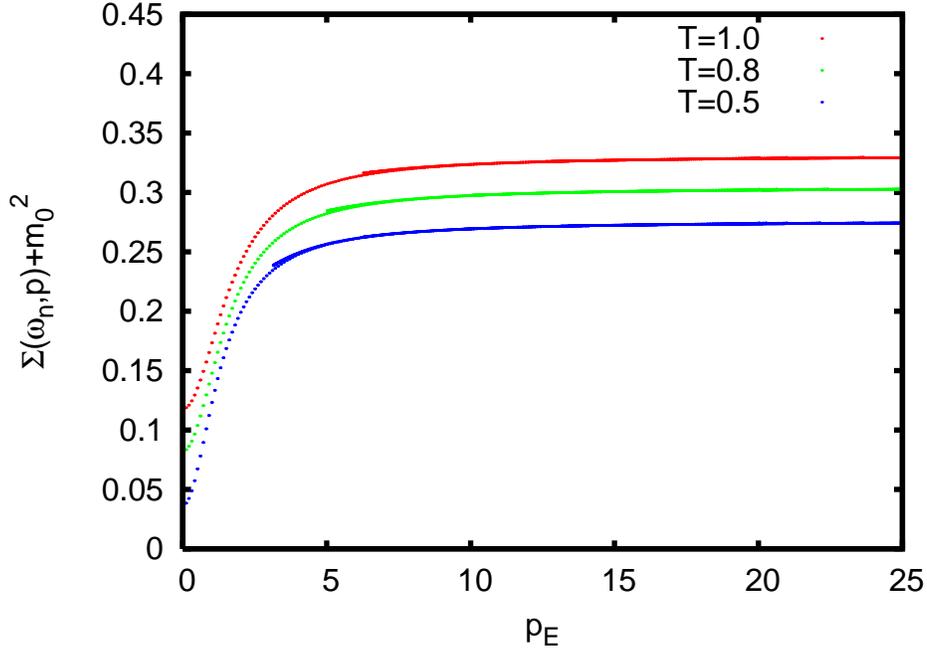}
  \caption{Momentum-dependent masses, cf. Eqs.~\eqref{eq:gap2PI},
    for different temperatures. $N=4$, $\lambda=0.5$.
  Here $p_E=\sqrt{\omega_n^2+p^2}$.}
  \label{fig:sigmas}
\end{figure}

\begin{figure}[htbp]
  \centering
  \epsfig{file=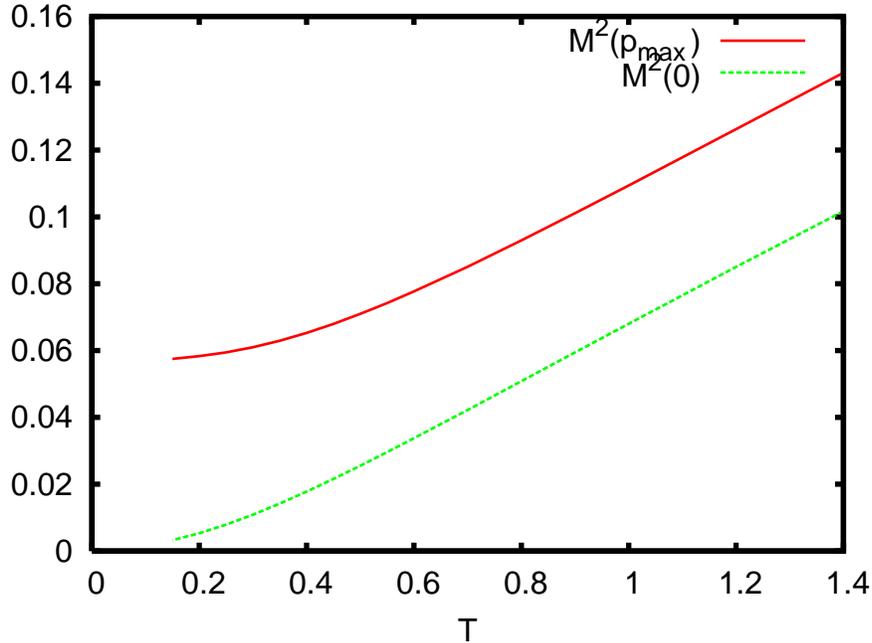}
  \caption{Temperature dependence of the effective 
    masses $m_0^2 + \Sigma(\omega,p)$
    at $\omega=p=0$ and in the asymptotic region
    ($p=p_{\rm max}$) for $\lambda=0.1$
    and $N=4$.}
  \label{fig:massesvsT}
\end{figure}

In Fig.~\ref{fig:sigmas} we display the spectral behavior of the
effective momentum-dependent masses 
$m_0^2+\Sigma(\omega_n,p)$, see Eqs.\eqref{eq:gap2PI},  
at the equilibrium point $\phi=0$, 
for various temperatures, $N=4$ and $\lambda=0.5$. 
We plot these quantities versus
the ``Euclidean momentum'' $p_E=\sqrt{\omega_n^2+p^2}$. The curves
for various Matsubara frequencies $\omega_n$ are close together,
implying an approximate rotation symmetry in the
Euclidean $(\omega,p)$ plane.  Whereas the large-momentum
behavior is determined by an effective constant mass
given by the tree level and bubble contributions, 
the effective masses at low momenta are considerably
smaller, so that the various loop integrals are
infrared-enhanced. 
The infrared enhancement can also be seen in  Fig. \ref{fig:massesvsT}
where we plot, for $\lambda=0.1$, the values of the effective masses
at the equilibrium point $\phi=0$. We display $m_0^2+\Sigma(0,0)$ and
$m_0^2+\Sigma(0,p_{\rm max})$, where $p_{\rm max}$ is our pragmatic
momentum cutoff. The values near $p=0$ are much smaller
than those at $p_{\rm max}$, which can be considered as the
asymptotic ones, see Fig. \ref{fig:sigmas}.

\section{Conclusions}
\label{sec:conclusions}
\setcounter{equation}{0}
We have analyzed here the $O(N)$ linear sigma model in the 2PI
formalism at next-to-leading order in a $1/N$ expansion.
In order to do so, we have solved the coupled system
of Euclidean Schwinger-Dyson equations for the $\sigma$ and $\pi$ 
propagators. 

We have found that within our range of parameters
the 2PI approximation does not display
spontaneous symmetry breaking, as to be expected in a
$1+1$ dimensional model. Furthermore, we find that the
thermal propagators are enhanced in the infrared region. Due
to momentum-dependent self-energy insertion the 
effective masses at low Euclidean momenta are smaller than those
at large momenta. 

We have compared the effective potential as a function of 
the external field $\phi$ to the ones obtained in
several other approximations.
The 1PI-NLO approximation is close to the 2PI-NLO one even for
small $N$, whereas the potential obtained in the 
2PPI-NLO approximation lies between those obtained in the
Hartree and in the 2PI approximations. All potentials
are above the large-$N$ effective potential and, of course, converge
to it at large $N$. The Hartree and 2PPI-NLO approximations
display spontaneous symmetry breaking, unphysical in $1+1$ dimensions.
This is clearly an artefact of these approximations.

It would be of interest to extend the analysis of the $O(N)$ 
linear sigma model in the 2PI-NLO approximation to nonequilibrium systems. 
For the $O(N)$ model out of equilibrium the 
2PPI approximation (to two loops in the effective action) 
displays a symmetric phase in the late time behavior
\cite{Heinen:2004ip}, in contrast to 
what we have found here in thermal equilibrium. 

Coming back to the simulations in $1+1$ dimensions for the case $N=1$ 
\cite{Cooper:2002qd,Baacke:2003qh} spontaneous symmetry breaking was
found in the late time behavior of the Hartree approximation,
whereas in both the 2PI and the 2PPI the mean field approaches
 the symmetric phase at late times. In contrast
to that, the 2PPI approximation in equilibrium displays spontaneous 
symmetry breaking, and so does the 2PI approximation both with 
correct $N=1$ combinatorics and with $1/N$ combinatorics 
\cite{Baacke:2004up}. This is due to the fact that these 
calculations do not encompass 
kink transitions which are of order $1/\lambda$. So there remain
some issues to be clarified.

In this work we were concerned mainly with the phase structure
of the model in the 2PI scheme at next-to-leading order
of the $1/N$ expansion.
It would of course be useful to extend this analysis to $3+1$ dimensions
using recently developed renormalizations techniques for
2PI-resummed perturbation theory 
\cite{vanHees:2001ik,vanHees:2001pf,vanHees:2002bv,
Blaizot:2003an,Cooper:2004rs}. Furthermore it 
would be interesting to compute various thermodynamic functions
and transport coefficients \cite{Aarts:2003bk,Aarts:2004sd}, 
in view of their relevance for heavy ion collisions.


\section*{Acknowledgments}
The authors take pleasure in thanking G. Aarts, J. Berges, S. Bors\'anyi,
A. Heinen, K. Rummukainen, M. Sall\'e and J. Serreau 
for useful discussions and comments on the manuscript.
S.M. was supported by DFG as a member of
\emph{Gra\-du\-ier\-ten\-kol\-leg~841}.


\begin{appendix}


\section{The 2PPI formalism to NLO of a $1/N$ expansion}
\label{sec:2PPI}
\setcounter{equation}{0}
In the 2PPI formalism \cite{Verschelde:1992bs,Coppens:1993zc} one 
introduces a quadratic source term
$K(x)\Phi^2(x)$ which, in contrast to the 2PI formalism, is local.
The Legendre transformed variables are then $\phi(x)$ and 
$\Delta(x)$. For a comparison between the 2PI and 2PPI formalism
see, e.g., Appendix~A of Ref.~\cite{Baacke:2002pi}. For a translation
invariant system $\phi$ and $\Delta$ are constants. The effective
action then becomes an ordinary function of these variables.
For the $O(N)$ system, the classical field is a vector
$\vec \Phi_{\rm cl}=\hat n \phi$.
As for the 2PI formalism we use a basis where the $\sigma$ field 
is in the direction $\hat n$ and the $N-1$ pion fields span 
the orthogonal directions. Then $\Delta$ is a $2\times2$ matrix
which becomes diagonal in this basis, with entries $\Delta_\sigma$ and
$\Delta_\pi$. In contrast to the 2PI formalism, the Green functions
take the simple form
\be\label{Green_2ppi}
\calg_j(p,\omega_n)=\frac{1}{p^2+\omega_n^2+\calm_j^2} 
\ee
with the effective masses
\begin{subequations}
\bea
\calm_\sigma^2&=&\lambda\left[3\phi^2-v^2+3\Delta_\sigma+(N-1)\Delta_\pi
\right]
\\
\calm_\pi^2&=&\lambda\left[\phi^2-v^2+\Delta_\sigma+(N+1)\Delta_\pi\right]
\pkt
\eea
\end{subequations}
Here the local mass insertions $\Delta_j$ are given by 
\be
\Delta_j=2\frac{\partial \Gamma^{\rm 2PPI}_{\rm q}}{\partial \calm_j^2}
\pkt
\ee
where $ \Gamma^{\rm 2PPI}_{\rm q}$ is the sum of all two particle
point irreducible graphs. These are all graphs that do not fall apart if two
internal lines meeting at one point are cut. They are computed with the
propagators defined in  Eq.~\eqn{Green_2ppi}.
As for the 2PI formalism we have considered the derivatives with
respect to $\calm_\pi^2$ as the derivative with respect to
the mass of one of the $(N-1)$ pion species, 
thus leaving out factors $(N-1)$
which anyway would cancel in the final formulae.

It is simpler to express the potential  in terms of 
the effective masses $\calm_j^2$ instead 
of the parameters
$\Delta_j$. Then these gap equations can be derived from the 
effective potential
\be
U_{\rm eff}(\phi,\mathcal{M}_\sigma^2,\mathcal{M}^2_\pi)
=N U_{\rm cl}(\phi,\mathcal{M}_\sigma^2,\mathcal{M}^2_\pi)
+U_{\rm q}(\phi,\mathcal{M}_\sigma^2,\mathcal{M}^2_\pi)
\pkt\ee
Here $U_{\rm cl}$ is the``classical'' potential 
\cite{Nemoto:1999qf,Baacke:2002pi}
\bea \label{Uclass_2PPI}
U_{\rm cl}&=&\frac{1}{2} \calm_\sigma^2\phi^2-\frac{\lambda}{2} \phi^4 -
\frac{1}{2\lambda (N+2)}v^2\left\{\calm_\sigma^2 +(N-1)\calm_\pi^2\right\}
\\ \nonumber
&&-\frac{1}{8\lambda(N+2)}\left[
(N+1)\calm_\sigma^4+3(N-1)\calm_\pi^4
-2(N-1)\calm_\sigma^2\calm_\pi^2+2 N \lambda^2 v^4\right]
\kma
\eea
which in this formulation already includes quantum parts through the
effective masses. In this sense the separation between
$U_{\rm cl}$ and the quantum part $U_{\rm q}$ is superficial. 
The basic contribution to $U_{\rm q}$ is the
one-loop contribution
\be
U_{\rm 1-loop}=U_{\rm 1-loop}^\sigma+(N-1)U_{\rm 1-loop}^\pi
\ee
with
\be \label{oneloop_2PPI1}
U_{\rm 1-loop}^j=\frac{1}{2}\tr \log \frac{\calg_j}{\calg_{j,0}}
\pkt\ee
which explicitly reads
\be \label{oneloop_2PPI2}
U^j_{\rm 1-loop}=\frac{1}{2} T \sum_n\int\frac{dp}{2\pi}
\ln \frac{p^2+\omega_n^2+\calm_j^2}{p^2+\omega_n^2+m_{j,0}^2}
\pkt\ee
In regularized and renormalized form it becomes
\bea\nonumber \label{oneloop_2PPI3}
U^j_{\rm 1-loop}&=&\frac{1}{2} T \sum_n\int\frac{dp}{2\pi}
\left\{\ln \frac{p^2+\omega_n^2+\calm_j^2}{p^2+\omega_n^2+m_{j,0}^2}\right.
\\&&\left.-\frac{\calm_j^2-m_{j,0}^2}{p^2+\omega_n^2+m_{j,0}^2}\right\}
+\frac{1}{2}(\calm_j^2-m_{j,0}^2)\calb_j
\pkt\eea
The regularized bubble integrals have been defined in Eq.~\eqn{bubble_reg}.
If we include just $U_{\rm 1-loop}$ we again obtain the Hartree
approximation with $\Delta_j=\calb_j$. 
The double-bubble diagrams are included here in the
$U_{\rm cl}$ via the effective masses.

\begin{figure}[htbp]
  \centering
  \epsfig{file=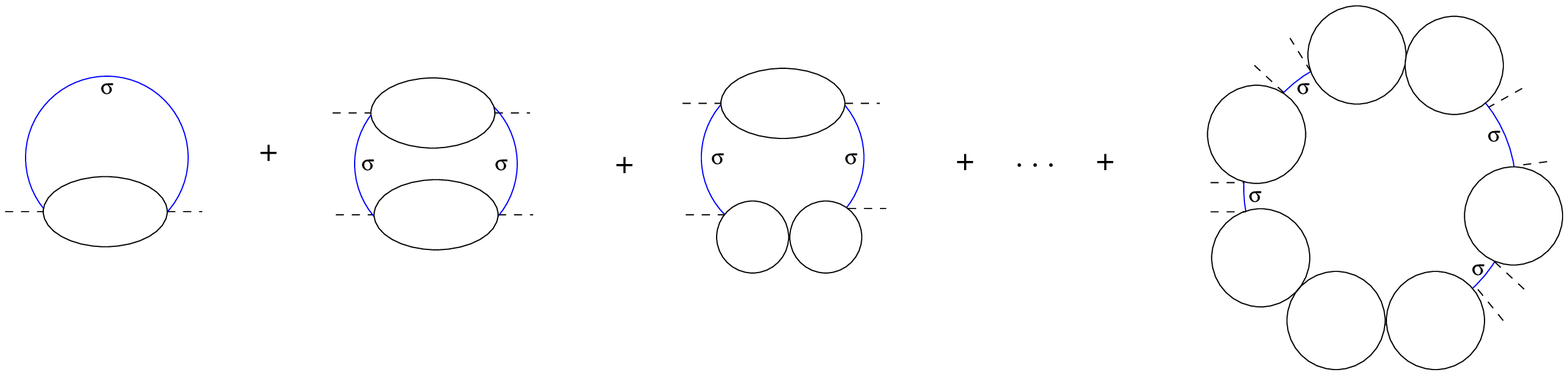,width=.9\textwidth}
  \caption{2PPI or 1PI "chain" contributions to the effective action.}
  \label{fig:chains}
\end{figure}

Going beyond the Hartree approximation in the 2PPI formalism 
in a strict $1/N$ expansion 
we only have to take into account ``necklaces''
and ``generalized sunsets'' and a 2PPI (but 2PR) combination of them.
The necklace $\caln$ takes the same form as for the 2PI
formalism, see Eq.~\eqn{defN} and Fig.\ref{fig:pearls}, 
where of course the fish diagram,
Eq.~\eqn{defF}, is
computed with the propagators of Eq.~\eqn{Green_2ppi}.
The generalized sunset contribution is replaced by 
a more complex set of graphs.
As discussed by Root~\cite{Root:1974zr}, at next-to-leading order
of $1/N$ the graphs with external lines have the form presented
in Fig.~\ref{fig:chains}, with alternating sigma propagators
and necklaces. As the resemblance to a sunset becomes
now very remote we refer to them as chain diagrams. 
These are summed up in the form
\begin{equation}
  \label{chain2PPI}
\calc=\frac{1}{2}T\sum_{m=-\infty}^\infty
\int\frac{dq}{2\pi}  
\ln\left[  1+\bfS(\omega_m,q)\right]
\end{equation}
where $\bfS(\omega_m,q)$ is the kernel of the
sunset diagram, Eq. \eqn{defS}:
\be \label{chaininsert2PPI}
\bfS(\omega_m,q)=-2\lambda\phi^2\frac{\lambda\calf(\omega_m,q)}{
1+\lambda\calf(\omega_m,q)}\calg_\sigma(\omega_m,q)
\ee
In the 2PI formalism the higher powers of insertions $\bfS$ are
included automatically, here the summation has to be done
explicitly. 

If the graphs defined in the previous paragraph are introduced  the
quantum part of the effective potential  takes the form
\be
U_{\rm q}=U_{\rm 1-loop}+\caln +\calc
\ee
and the insertions $\Delta_j$ in the gap
equations are given by
\be
\Delta_j=\calb_j+2\frac{\partial \caln}{\partial \calm^2_j}+
2\frac{\partial \calc}{\partial \calm^2_j}
\pkt\ee
Explicitly the new contributions are given  by
\be \label{dndmj2}
\frac{\partial \caln}{\partial \calm^2_j}=\frac{1}{2}
\frac{\lambda}{N}T\sum_n\int\frac{dp}{2\pi}\frac{-\lambda\calf(q,\omega_n)}
{1+\lambda\calf(q,\omega_n)}
\frac{\partial\calf_j}{\partial \calm_j^2}(q,\omega_n)
\ee
and
\bea\nonumber
\frac{\partial \calc}{\partial \calm^2_j}
&=&-\lambda\phi^2 T\sum_n\int\frac{dq}{2\pi}\frac{1}{1+\bfS(\omega_m,q)}
\left\{\frac{\lambda}{N}\frac{1}{[1+\lambda \calf(\omega_m,q)]^2}
\frac{\partial \calf_j(\omega_m,q)}{\partial \calm_j^2}\calg_\sigma(\omega_n,q)
\right.
\\&&\left.
-\delta_{j\sigma}\frac{\lambda\calf(\omega_m,q)}
{1+\lambda\calf(\omega_m,q)}
\calg_\sigma^2(\omega_m,q)
\right\}
\eea
Finally $\calf=(\calf_\sigma+(N-1)\calf_\pi)/N$ as before, and 
\be
\frac{\partial\calf_j}{\partial \calm_j^2}(p,\omega_n)
=-2 T\sum_M\int\frac{dq}{2\pi}
\frac{1}{(q^2+\omega_m^2+\calm_j^2)\,\left[(p-q)^2+(\omega_n-\omega_m)^2
+\calm_j^2 \right]^2}
\pkt\ee
The partial derivative of $V_{\rm eff}$ with respect to $\phi$, divided
by $N\phi$, in analogy to Eq. \eqn{defR} is given by
\be
\calr(\phi) = \calm_\sigma^2-2\lambda \phi^2+
\frac{1}{N\phi}\frac{\delta \calc}{\delta \phi}
\ee
with
\be
\frac{1}{N\phi}\frac{\partial \calc}{\partial \phi}
=-2\frac{\lambda}{N} T\sum_n\int\frac{dq}{2\pi}\frac{1}{1+\bfS(\omega_m,q)}
\frac{\calf(\omega_m,q)}
{1+\lambda\calf(\omega_m,q)}
\calg_\sigma(\omega_m,q)
\ee


\section{The 2PPI formalism to NLO of a $1/N$ expansion}
\label{sec:1PI1/N}
\setcounter{equation}{0}
In a seminal article \cite{Root:1974zr} Root has discusssed the extension
to the next-to-leading order in a $1/N$ expansion within the 1PI
formalism. He introduces an auxiliary field $\chi$ which in the language
used throughout our work can be identified with a self-consistent
pion mass $\calm_\pi^2$. The classical potential then takes the form
\be
U_{\rm cl}(\phi,\calm_\pi^2)=N\left(\frac{1}{2}\calm_\pi^2(\phi^2-v^2)
-\frac{\calm_\pi^4}{4\lambda}\right)
\kma\ee
it can be obtained by taking the large-$N$ limit of the analogous potential
of the 2PPI formalism, Eq. \eqn{Uclass_2PPI},
when the mass of the $\sigma$ propagator is given by
\be \label{msigma_root}
\calm_\sigma^2=\calm_\pi^2+2\lambda\phi^2
\pkt
\ee  
To leading order in $1/N$ this would follow in the 2PPI formalism
when taking only the pions into account in the one-loop terms.
In Root's work this  relation is kept fixed for all orders of $1/N$, 
however. In a strict
$1/N$ counting one may indeed neglect corrections of NLO to the propagator, 
as actually suggested in Root's
work by using the leading order gap equation.
As in the 2PPI formalism the propagators take the tree level form
of Eq.~\eqn{Green_2ppi}. 

In the following we will rewrite the various terms included by Root
in a way that makes the correspondence to the other formalisms
more transparent. Including the next-to-leading order corrections
the one-loop term takes the form of 
Eqs.~\eqn{oneloop_2PPI1}-\eqn{oneloop_2PPI3}, 
with $\calm_\sigma^2$ given by Eq.
\eqn{msigma_root}. As in the 2PPI formalism the next terms are
the necklace  and the chain diagrams. In the necklace
part only the pion fish diagram is taken into account. The definition
of the necklace contribution is then identical to the one in the
2PI formalism, Eq. \eqn{defN} with the modification that
$\calf\equiv\calf_\pi$ and that the propagators in the fish graph,
Eq. \eqn{defF} are given by Eq. \eqn{Green_2ppi}.

The chain diagram $\calc$ takes the same form as in the 2PPI formalism,
see Eqs. \eqn{chain2PPI} and \eqn{chaininsert2PPI}, of course now
with $\calf=\calf_\pi$. In taking the derivatives
the sigma propagator is now considered as a function of $\calm_\pi^2$ and 
$\phi$. The modifications with respect to the 2PPI formalism are obvious.
We have
\bea\nonumber
\frac{\partial \calc}{\partial \calm^2_\pi}
&=&-\lambda\phi^2 T\sum_n\int\frac{dq}{2\pi}\frac{1}{1+\bfS(\omega_m,q)}
\left\{\frac{\lambda}{N}\frac{1}{[1+\lambda \calf(\omega_m,q)]^2}
\frac{\partial \calf_j(\omega_m,q)}{\partial \calm_\pi^2}
\calg_\sigma(\omega_n,q)
\right.
\\&&\left.
-\frac{\lambda\calf(\omega_m,q)}
{1+\lambda\calf(\omega_m,q)}
\calg_\sigma^2(\omega_m,q)
\right\}
\eea
and
\be
\frac{1}{N\phi}\frac{\partial \calc}{\partial \phi}
=-2\frac{\lambda}{N} T\sum_n\int\frac{dq}{2\pi}\frac{1}{1+\bfS(\omega_m,q)}
\frac{\calf(\omega_m,q)}
{1+\lambda\calf(\omega_m,q)}
\calg_\sigma(\omega_m,q)\left[1-2\lambda\phi^2\calg_\sigma(\omega_m,q)\right]
\pkt
\ee
Finally the gap equation is
\be
\calm_\pi^2=\lambda\left(\phi^2-v^2
+2\frac{\partial \caln}{\partial \calm^2_\pi}
+2\frac{\partial \calc}{\partial \calm^2_\pi}\right)
\ee 
where $\partial \caln / \partial \calm^2_\pi$
is given by Eq. \eqn{dndmj2}, and the quantity $\calr$ is given by
\be
\calr(\phi) = \calm_\sigma^2-2\lambda \phi^2+
\frac{1}{N\phi}\frac{\delta \calc}{\delta \phi}
\pkt
\ee

\end{appendix}

\bibliography{refs}
\bibliographystyle{h-physrev4}

\end{document}